\documentclass[letterpaper,twocolumn,american,showpacs,pra,aps,superscriptaddress]{revtex4}
\usepackage[latin1]{inputenc}
\usepackage{bm}
\usepackage{multirow,amssymb,amsbsy,amsmath}
\usepackage{stmaryrd}
\usepackage{graphicx}
\usepackage{color}
\makeatletter
\usepackage{pifont}
\makeatother
\DeclareMathOperator{\Tr}{Tr}
\newcommand{\ket}[1]{|{#1}\rangle}

\begin{document}
%%%%%%%%%%%%%%%%%%%%%%%%%%%%%%%%%%%%%%%%%%%%%%%%%%%%%%%%%%%%%%%%%%%%%%%%%

\title{Non-Markovian dissipative dynamics of two coupled qubits in independent reservoirs:\\a comparison between exact solutions and master equation approaches}

\author{E. Ferraro}
\email{ferraro@fisica.unipa.it}\affiliation{Dipartimento di Scienze
Fisiche ed Astronomiche, Universit\`a di Palermo, via Archirafi 36,
90123 Palermo, Italy}

\author{M. Scala}
\email{matteo.scala@fisica.unipa.it}\affiliation{Dipartimento di
Scienze Fisiche ed Astronomiche, Universit\`a di Palermo, via
Archirafi 36, 90123 Palermo, Italy}\affiliation{Department of
Physics, Sofia University, James Bourchier 5 blvd., 1164 Sofia,
Bulgaria }

\author{R. Migliore}
\affiliation{CNR-INFM, Research Unit CNISM of Palermo, via Archirafi
36, 90123 Palermo, Italy}

\author{A. Napoli}
\affiliation{Dipartimento di Scienze Fisiche ed Astronomiche,
Universit\`a di Palermo, via Archirafi 36, 90123 Palermo, Italy}

\date{\today}

\begin{abstract}
The reduced dynamics of two interacting qubits coupled to two
independent bosonic baths is investigated. The one-excitation
dynamics is derived and compared with that based on the resolution
of appropriate non-Markovian master equations. The Nakajima-Zwanzig
and the time-convolutionless projection operator techniques are
exploited to provide a description of the non-Markovian features of
the dynamics of the two-qubits system. The validity of such
approximate methods and their range of validity in correspondence to
different choices of the parameters describing the system are
brought to light.
\end{abstract}


\maketitle

\section{Introduction}

Despite its simplicity a two-state system is of great significance
being it exploitable to effectively describe many real situations.
The theoretical analysis as well as the practical implementation of
interacting or not two-level systems thus represents a central topic
in several branches of modern physics ranging from high energy to
nuclear and condensed matter physics \cite{polyakov,belitsky}.
During the last decade the interest towards two-level systems has
further been stimulated by the fact that a qubit represents the
basic element in the context of the new applicative area of quantum
information and communication. It has been for example shown that a
fundamental quantum gate like the C-NOT gate can be implemented
using dipole-dipole interacting quantum dots modeled as two qubits
\cite{baranco}. Moreover due to the development of new technologies,
today there are several possible routes to the creation of what
might be termed quantum bit, each based on a different physical
system. These include quantum optics, microscopic quantum objects
(electrons, ions, atoms) in traps, quantum dots and quantum circuits
\cite{amico,6,7,8,9,10,bellomo,garraway1,garraway2,garraway4}. In
describing real systems however it is mandatory to take into account
the effects stemming from the presence of the surroundings. Thus the
dissipative dynamics of two-level systems has been the subject of
numerous papers appeared in literature in the last decades
\cite{agarwal,sinaysky,quiroga,maniscalco,Mazzola,yu,scala}.
Generally speaking the research has been developed assuming a
Markovian environment \cite{agarwal,sinaysky}. But memory effects
are in general present and could affect quantitatively and
qualitatively the dynamics of the small system. Unfortunately, there
are no fully systematic investigations of non-Markovian
environments. Projection operator techniques, such as the
time-convolutionless (TCL) \cite{Chaturvedi} and the
Nakajima-Zwanzig (NZ) \cite{Nakajima,Zwanzig} approaches, are in
general exploited in order to perform a description of the
non-Markovian features of the dynamics of open systems. On the one
hand, the NZ provides a generalized master equation in which the
time derivative of the density operator is connected to the past
history of the state through the convolution of the density operator
and an appropriate integral kernel. On the other hand the TCL
approach provides a generalized master equation which is local in
time. Intuitively, one might argue that the NZ should work better
than the TCL approach in describing the memory effect, since it
explicitly takes into account the past history of the open system.
Anyway there are examples in which the exact dynamics of the open
system can be described by means of a master equation which is local
in time, as in the well known case of the Hu-Paz-Zhang generalized
master equation for the non-Markovian theory of quantum Brownian
motion \cite{hu,petruccionebook}.

In general it is not easy to establish whether one method is better
than the other one. In fact, the performance of these perturbation
schemes strongly depends on the details of the system under
investigation. In this paper we consider two interacting two-level
systems, each coupled to its own bosonic bath, exactly solving their
dynamics in a one excitation subspace. The knowledge of the exact
dynamics is exploited to test perturbative approaches based on TCL
and NZ techniques. We show that, counterintuitively, the TCL
approach works better than the NZ one, since the latter approach
does not guarantee the positivity of the density matrix when the
correlations inside the reservoir become moderately strong. On the
contrary the TCL approach describes all the qualitative features of
non-Markovian dynamics for a wider range of values of reservoir
memory time.

The paper is organized as follows. In Sec. II we derive the exact
equations governing the evolution of the two-qubit system coupled to
two independent reservoirs, in the case of one initial excitation,
and find their exact solution. In Sec. III the derivation of the
second order NZ equation is presented and the features of the second
order TCL, derived in Ref.\cite{Ferraro}, are recalled. An extensive
comparison among the exact, the NZ and the TCL is presented in Sec.
IV, while in Sec. V some conclusive remarks are given.

\section{Exact dynamics}

\subsection{The model}
The physical system on which we focus our attention is composed by
two interacting two-level systems. Each qubit is moreover coupled to
an external environment modeled as a bosonic bath \cite{quiroga}.
Assuming $\hbar=1$, the Hamiltonian model describing the total
system can be written in the following form
\begin{equation}\label{Htot}
H=H_0+H_I.
\end{equation}
Here
\begin{eqnarray}\label{H0}
H_0=\frac{\omega_0}{2}\sigma_z^{(1)}+\frac{\omega_0}{2}\sigma_z^{(2)}+
\sum_{j=1,2}\sum_k\omega_k^{(j)}b_k^{(j)\dagger}b_k^{(j)}
\end{eqnarray}
is the unperturbed part containing the free Hamiltonian of the two
qubits as well as that of the two independent environments. The
transition frequency of the two two-level systems, supposed
coincident for simplicity, is indicated by $\omega_0$ whereas
$\sigma_z^{(j)}$ ($j = 1,2$) denotes the Pauli operator describing
the $j$-th subsystem. The two independent bosonic baths are
characterized by proper frequencies $\omega_k^{(j)}$,
$b_k^{(j)\dagger}$ and $b_k^{(j)}$ being correspondingly the
creation and annihilation bosonic operators.

The interaction term
\begin{equation}\label{Hint}
\begin{split}
H_I=&\Omega(\sigma_+^{(1)}\sigma_-^{(2)}+\sigma_-^{(1)}\sigma_+^{(2)})+\\
&+\sum_{j=1,2}\left(\sigma_+^{(j)}\sum_kg_k^{(j)}b_k^{(j)}+\sigma_-^{(j)}\sum_kg_k^{(j)\ast}b_k^{(j)\dagger}\right)
\end{split}
\end{equation}
includes both the direct interaction between the two qubits,
characterized by the coupling constant $\Omega$, and the interaction
between each qubit and its respective bosonic bath, with coupling
constants $g_k^{(j)}$. In eq.(\ref{Hint}) $\sigma_{\pm}^{(j)}\equiv
\frac{1}{2}(\sigma_{x}^{(j)} \pm i \sigma_{y}^{(j)})$ are, as usual,
the lowering and raising Pauli operators.

It is worth underlining that the Hamiltonian model \eqref{Htot} is
quite versatile in the sense that it can be successfully adopted to
describe many different physical systems. In the framework of cavity
QED \cite{bellomo} or circuit QED \cite{6,7,8,9,10} this model can
be indeed exploited for the description of two atoms in spatially
separated cavities as well as of two far enough Josephson charge,
flux or phase qubits so that it is reasonable to assume that they
interact with two different electromagnetic environments. Model
\eqref{Htot} in addition allows to study the influence of spurious
microwave resonators within Josephson tunnel junctions on the
coherent dynamics of a phase qubit \cite{martinis}. In this case the
environment coupled to the spurious resonator (modeled as a two
state system) is a phononic bath.

Very recently the Markovian dynamics stemming from Hamiltonian
\eqref{Htot} has been analyzed \cite{sinaysky}. In
Ref.\cite{Ferraro} instead the TCL approach has been exploited in
order to investigate on the non-Markovian regime. In this paper we
exactly solve the time-dependent Schr\"{o}dinger equation confining
ourselves to the one excitation subspace. At the same time we adopt
the NZ technique to derive a non-Markovian master equation for the
reduced density operator of the two coupled qubits. Having at our
disposal both the exact dynamics and the approximate master
equations, a comparison may be done in order to test the
effectiveness of the perturbative approaches. From now on, we work
in the interaction picture defined by $H_0$ in which the interaction
Hamiltonian reads
\begin{equation}\label{Hint2}
H_I(t)=H_I^{(s)}+H_I^{(D_1)}+H_I^{(D_2)}
\end{equation}
with
\begin{equation}
H_I^{(s)}=\Omega(\sigma_+^{(1)}\sigma_-^{(2)}+\sigma_-^{(1)}\sigma_+^{(2)})
\end{equation}
and
\begin{equation}\begin{split}
H_I^{(D_j)}&=\sigma_+^{(j)}\sum_kg_k^{(j)}b_k^{(j)}e^{i(\omega_0-\omega_k^{(j)})t}+\\
&+\sigma_-^{(j)}\sum_kg_k^{(j)\ast}b_k^{(j)\dagger}e^{-i(\omega_0-\omega_k^{(j)})t}.
\end{split}\end{equation}

\subsection{One excitation time evolution}
Let us begin by looking at the exact solution of the time-dependent
Schr\"{o}dinger equation. It is easy to verify that the number
operator
\begin{eqnarray}
\hat{N}=\sum_{j=1,2}\sigma_+^{(j)}\sigma_-^{(j)}+\sum_{j=1,2}\sum_{k}b_k^{(j)\dagger}b_k^{(j)}
\end{eqnarray}
is a constant of motion. This in particular means that, starting
from an eigenstate of $\hat{N}$, the system evolves remaining in the
subspace correspondent to the same eigenvalue $n$ of $\hat{N}$. In
what follows we consider the dynamics of the system in the subspace
with one excitation, that means $n=1$. To this end let us suppose to
prepare at $t=0$ the two qubits in a linear superposition of states
with one excitation and  both the baths in the vacuum state denoted
by $|0_k^{(j)}\rangle$ ($j=1,2$)
\begin{equation}
|\psi(0)\rangle=\left(a(0)|10\rangle+b(0)|01\rangle\right)|0_k^{(1)}0_k^{(2)}\rangle,
\end{equation}
with $|a(0)|^2+|b(0)|^2=1$. Since $[H,\hat{N}]=0$, at a generic time
instant $t$ we may write
\begin{equation}\label{psi}\begin{split}
&|\psi(t)\rangle=\left(a(t)|10\rangle+b(t)|01\rangle\right)|0_k^{(1)}0_k^{(2)}\rangle+\\
&+|00\rangle\left(\sum_kc_k^{(1)}(t)|1_k^{(1)}0_k^{(2)}\rangle+\sum_kc_k^{(2)}(t)|0_k^{(1)}1_k^{(2)}\rangle\right),
\end{split}\end{equation}
where $|1_k^{(j)}\rangle$ denotes a state of the $j-th$ bath
($j=1,2$) with one excitation in the mode $k$ and the probability
amplitudes $a(t)$, $b(t)$ and $c_k^{(j)}$ ($j=1,2$) are solutions of
the following system of differential equations
\begin{equation}\label{diff_eq_ampl}
\left\{\begin{array}{c} \dot{a}(t)=-i\left(\Omega\,
b(t)+\sum_kc_k^{(1)}(t)g_k^{(1)}e^{i(\omega_0-\omega_k^{(1)})t}\right)\\
\dot{b}(t)=-i\left(\Omega\,
a(t)+\sum_kc_k^{(2)}(t)g_k^{(2)}e^{i(\omega_0-\omega_k^{(2)})t}\right)\\
\dot{c}_k^{(1)}(t)=-i\,a(t)g_k^{(1)\ast}e^{-i(\omega_0-\omega_k^{(1)})t}\\
\dot{c}_k^{(2)}(t)=-i\,b(t)g_k^{(2)\ast}e^{-i(\omega_0-\omega_k^{(2)})t}.\\
\end{array}\right.
\end{equation}

From eqs.(\ref{diff_eq_ampl}) it is easy to verify that the
amplitudes $c_k^{(j)}(t)$ formally evolve as follows:
\begin{equation}
c_k^{(1)}(t)=-ig_k^{(1)\ast}\int_0^ta(t')e^{-i(\omega_0-\omega_k^{(1)})t'}\,dt'
\end{equation}
\begin{equation}
c_k^{(2)}(t)=-ig_k^{(2)\ast}\int_0^tb(t')e^{-i(\omega_0-\omega_k^{(2)})t'}\,dt'.
\end{equation}
Inserting these formal solutions in the equations for $a(t)$ and
$b(t)$ we achieve
\begin{equation} \label{system}
\left\{\begin{array}{c} \dot{a}(t)=-i\Omega\,
b(t)-i\int_0^ta(t')f_1(t-t')\,dt'\\
\dot{b}(t)=-i\Omega\,
a(t)-i\int_0^tb(t')f_2(t-t')\,dt'\\
\end{array}\right.
\end{equation}
where the kernel $f_j(t-t')$ is given by the correlation function
defined as
\begin{equation}\label{fj}
f_j(t-t')=\sum_k|g_k^{(j)}|^2e^{i(\omega_0-\omega_k^{(j)})(t-t')}
\end{equation}
that in the continuum limit becomes
\begin{equation}\label{fjcont}
f_j(t-t')=\int_0^{+\infty}d\omega\,J_j(\omega)e^{i(\omega_0-\omega)(t-t')},
\end{equation}
$J_j(\omega)$ being the spectral density of the $j-th$ bath.

Making use of the Laplace transform, the system \eqref{system}
becomes
\begin{equation}\label{systemLaplace}
\left\{\begin{array}{c} s\,\tilde{a}(s)-a(0)=-i\Omega\,\tilde{b}(s)-\tilde{a}(s)\tilde{f}_1(s)\\
s\,\tilde{b}(s)-b(0)=-i\Omega\,\tilde{a}(s)-\tilde{b}(s)\tilde{f}_2(s),
\end{array}\right.
\end{equation}
where $\tilde{a}(s)$, $\tilde{b}(s)$ and $\tilde{f}_j(s)$ denote the
Laplace transforms of $a(t)$, $b(t)$ and $f_j(t-t')$ respectively.
It is thus immediate to obtain
\begin{equation}\label{a}
\tilde{a}(s)=\frac{a(0)(s+\tilde{f}_2(s))-i\Omega
b(0)}{(s+\tilde{f}_1(s))(s+\tilde{f}_2(s))+\Omega^2}
\end{equation}
\begin{equation}\label{b}
\tilde{b}(s)=\frac{b(0)(s+\tilde{f}_1(s))-i\Omega
a(0)}{(s+\tilde{f}_1(s))(s+\tilde{f}_2(s))+\Omega^2}.
\end{equation}
Once fixed the spectral densities for both baths $J_1(\omega)$ and
$J_2(\omega)$, it is quite easy to obtain the time behavior of
$a(t)$, $b(t)$ and $c_k^{(j)}(t)$, simply antitransforming the
amplitudes $\tilde{a}(s)$ and $\tilde{b}(s)$ given by eqs.\eqref{a}
and \eqref{b}. The results thus obtained will be discussed in Sec.
IV.

\section{Nakajima-Zwanzig master equation}
In this section we will apply the projection operator techniques in
order to derive a non-Markovian master equation for the reduced
density matrix $\rho_S(t)$ of the two qubits \cite{petruccionebook}.
To this end, it is convenient to introduce a super-operator
according to
\begin{equation}
\rho\rightarrow\mathcal{P}\rho=\Tr_B{\rho}\otimes\rho_B\equiv\rho_S\otimes\rho_B,
\end{equation}
where $\rho_B$ is the density matrix of the environment. The
super-operator $\mathcal{P}$ projects any state of the total system
$\rho$ onto its relevant part $\mathcal{P}\rho$, expressing formally
the elimination of the irrelevant degrees of freedom from the full
dynamical description of the model under scrutiny. Following the NZ
approach we get an integro-differential equation
\begin{equation}\label{maste}
\frac{d}{dt}\mathcal{P}\rho(t)=\int_0^tdt'\,\mathcal{K}(t,t')\mathcal{P}\rho(t')
\end{equation}
describing the reduced dynamics of the system. Here the memory
kernel $\mathcal{K}(t,t')$ is a super-operator in the relevant
subspace. In order to discuss the reduced dynamics we perform a
perturbation expansions of $\mathcal{K}(t,t')$ with respect to the
strength of the interaction Hamiltonian \cite{petruccionebook}. If
we restrict ourselves to the second order, the two relevant terms of
$\mathcal{K}(t,t')$ can be written down as
\begin{equation}
\mathcal{K}_1(t,t')=\mathcal{P}\mathcal{L}(t)\mathcal{P}
\end{equation}
and
\begin{equation}
\mathcal{K}_2(t,t')=\mathcal{P}\mathcal{L}(t)\mathcal{Q}\mathcal{L}(t')\mathcal{P},
\end{equation}
where $\mathcal{Q}\equiv I-\mathcal{P}$ and $\mathcal{L}$ is the
Liouville super-operator defined by
$\mathcal{L}(t)\rho(t)\equiv-i[H_I(t),\rho(t)]$. Starting from
eq.\eqref{maste} it is possible to demonstrate that the second order
non-Markovian master equation assumes the following form
\begin{equation}\label{master0}
\begin{split}
\frac{d}{dt}\rho_S(t)=&-i[H_I^{(s)},\rho_S(0)]+\\
&-\int_0^tdt'\,\Tr_B\left\{[H_I(t),[H_I(t'),\rho_S(t')\otimes\rho_B]]\right\},
\end{split}
\end{equation}
that, contrarily to the TCL master equation, is a non-local
evolution equation. After some manipulations it is possible to
recast eq.\eqref{master0} in the more compact form
\begin{equation}\label{master}
\begin{split}
\frac{d}{dt}\rho_S(t)=&-i[H_I^{(s)},\rho_S(0)]+\mathcal{L}^{(s)}\rho_S(t)+\\
&+\mathcal{L}^{(D_1)}\rho_S(t)+\mathcal{L}^{(D_2)}\rho_S(t),
\end{split}
\end{equation}
where the dissipators are expressed by
\begin{equation}
\mathcal{L}^{(s)}\rho_S(t)=-\int_0^tdt'\,[H_I^{(s)}(t),[H_I^{(s)}(t'),\rho_S(t')]]
\end{equation}
and
\begin{equation}\label{dj}
\begin{split}
\mathcal{L}^{(D_j)}\rho_S(t)=\int_0^t&dt'\,\left\{g_j(t-t')[\sigma_-^{(j)}\rho(t'),\sigma_+^{(j)}]+\right.\\
&\left.+g_j^{\ast}(t-t')[\sigma_-^{(j)},\rho_S(t')\sigma_+^{(j)}]\right\}.
\end{split}
\end{equation}
In eq.\eqref{dj} $g_j(t-t')$ is the correlation function, that in
correspondence to a thermal bath with $T=0$ coincides with
$f_j(t-t')$ defined in eq.\eqref{fj}. Once again, in order to solve
the master equation \eqref{master} we can exploit the Laplace
transform. The results we have obtained are reported in the next
section where they are compared with the exact dynamics as well as
with the results obtained exploiting the TCL approach
\cite{Ferraro}. The master equation solved in Ref.\cite{Ferraro} is
a time-local differential equation and it can be obtained from
eq.\eqref{master0} by replacing $\rho(t')$ with $\rho(t)$.

\section{Comparison between exact and approximate solutions}
Exploiting the results obtained in the previous sections we now
analyze the time behavior of the two qubits comparing in particular
the exact dynamics with the ones stemming from the NZ and TCL
approaches. Our aim is to highlight the performances of two
approximate approaches and to point out their range of validity. As
underlined in the introduction there is not indeed a general theory
that predict which one of the two methods is to be preferred and
generally speaking their range of validity strongly depends on the
specific features of the system, namely the interaction hamiltonian,
the interaction time, the environmental state and the spectral
density.

In what follows we assume that each qubit interacts resonantly with
a reservoir with Lorentzian spectral density
\begin{equation}
J_1(\omega)=J_2(\omega)\equiv
J(\omega)=\frac{1}{2\pi}\frac{\gamma\lambda^2}{(\omega_0-\omega)^2+\lambda^2},
\end{equation}
where $\gamma$ is a parameter which in the Markovian limit coincides
with the system decay rate, and $\lambda$ is the reservoir
bandwidth. This is for instance the case of two atoms interacting
each of them with their own cavity field in presence of cavity
losses \cite{garraway1}. Thus the two correlation functions
$f_1(t-t')$ and $f_2(t-t')$ defined by eq.\eqref{fjcont} coincide
and are given by
\begin{equation}\label{corr_lorentz}
f_1(t-t')=f_2(t-t')\equiv f(t-t')=\frac{1}{2}\gamma\lambda
e^{-\lambda|t-t'|}.
\end{equation}
From the latter equation, it is clear that the bandwidth $\lambda$
plays the role of the inverse of the reservoir memory time.

The system dynamics will be analyzed considering the two baths in a
thermal state at $T=0$ whereas the two qubits are supposed at $t=0$
in the Bell state
$|\psi(0)\rangle=\frac{1}{\sqrt{2}}\left(|10\rangle-|01\rangle\right)$.
Let us concentrate on the population $P_{10}(t)$, that is on the
probability of finding the two qubits in the state $|10\rangle$. To
test the validity of the approximate approaches we explore three
different regimes varying the width of the Lorentzian spectral
density $\lambda$. This investigation will allow us to assess in
which cases the solutions of the master equations are efficient in
the description of the true dynamics of the system.

%%%%---------------------------- FIGURE 1 --------------------%%

\begin{figure}
 \begin{center}
     \includegraphics[width=0.4\textwidth]{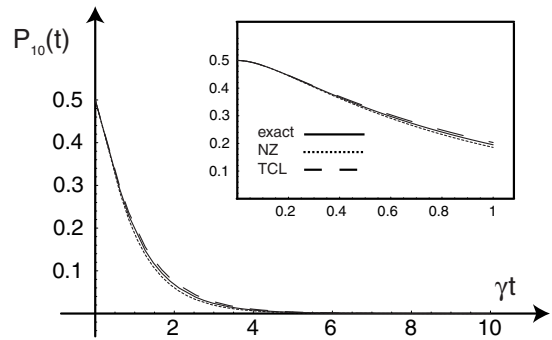}
    \caption{Time evolution of the population of the state $\ket{10}$
    for a system initially prepared in the Bell state
    $\frac{1}{\sqrt{2}}\left(\ket{10}-\ket{01}\right)$.
    The width of the Lorentzian spectral density is $\lambda=10\gamma$, the strength of the coupling constant
    between the two qubits is $\Omega=0.001\gamma$.}\label{figura1}
 \end{center}
\end{figure}

%%%%------------------------------------------------%%

Figure \eqref{figura1} shows a comparison among the exact, the TCL
and the NZ solutions in the case of large reservoir bandwidth
$\lambda=10\gamma$. The plot is done against the dimensionless
variable $\gamma t$ and the coupling constant between the qubits is
fixed to $\Omega=0.001\gamma$. We can clearly appreciate the perfect
agreement between the exact analytical solution and the approximate
ones for the short time behavior but also for long interaction
times. In this case the two approaches TCL and NZ both provide a
very good description of the dynamics and we may conclude that there
is no way to establish if one method is to be preferable with
respect to the other, they indeed give the same results. However in
such cases the TCL master equation might be preferred since it
involves a time local first order differential equation and
therefore it is easier to solve. In the inset of fig.\eqref{figura1}
the short time behavior is shown. It is interesting to underline the
initially quadratic behavior that witnesses the non-Markovian
features of the dynamics.

%%%%---------------------------- FIGURE 2 --------------------%%

\begin{figure}
 \begin{center}
     \includegraphics[width=0.4\textwidth]{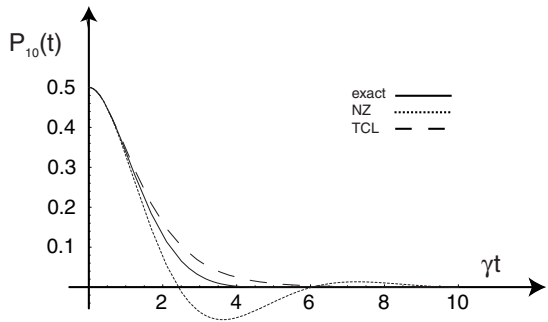}
    \caption{Time evolution of the population of the state $\ket{10}$
    for a system initially prepared in the Bell state
    $\frac{1}{\sqrt{2}}\left(\ket{10}-\ket{01}\right)$.
    The width of the Lorentzian spectral density is $\lambda=\gamma$, the strength of the coupling constant
    between the two qubits is $\Omega=0.001\gamma$.}\label{figura2}
 \end{center}
\end{figure}

%%%%------------------------------------------------%%

In fig.\eqref{figura2} the same quantity is reported choosing
$\lambda=\gamma$. Despite the good agreement for the short time
dynamics, we observe significant deviations when time increases. In
particular concerning the long time behavior, the NZ equation leads
to a very bad approximation. For times longer than some critical
values the solution for the population $P_{10}(t)$ cannot represent
a true diagonal element of a density matrix anymore, since it indeed
assumes negative values. We may conclude that for this range of
parameters the TCL solution gives a better description of the
dynamics since it reproduces all the qualitative features of the
exact solution.

%%%%---------------------------- FIGURE 3 --------------------%%

\begin{figure}
 \begin{center}
     \includegraphics[width=0.4\textwidth]{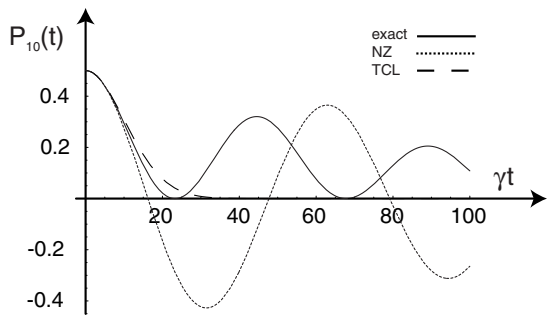}
    \caption{Time evolution of the population of the state $\ket{10}$
    for a system initially prepared in the Bell state
    $\frac{1}{\sqrt{2}}\left(\ket{10}-\ket{01}\right)$.
    The width of the Lorentzian spectral density is $\lambda=0.01\gamma$, the strength of the coupling constant
    between the two qubits is $\Omega=0.001\gamma$.}\label{figura3}
 \end{center}
\end{figure}

%%%%------------------------------------------------%%

Finally in fig.\eqref{figura3} we examine the regime
$\lambda=0.01\gamma$ which, according to eq.\eqref{corr_lorentz},
corresponds to very strong reservoir correlations and very long
memory time. We observe once again a perfect agreement among all the
three approaches in the short time behavior but in this case we
assist at a failure of the TCL approach too. The solution of the TCL
master equation (dashed line) doesn't succeed to follow the Rabi's
oscillations witnessed by the exact dynamics (solid line). The NZ
approaches presents the same problem of not conserving the
positivity of the density matrix as in the previous case. Thus, in
this case, both the perturbative approaches are not suitable to
describe the dynamics of the system. It is interesting to observe
that the oscillations appearing in the exact evolution of the
populations $P_{10}(t)$ are not due to the spin-spin interaction
constant $\Omega$, which we have taken small on purpose. The
oscillations are instead due to the fact that, in this case, the
Lorentzian peak is so narrow to make the environment equivalent to a
cavity with losses, as one might verify for example by using the
pseudomode approach in Refs.\cite{Mazzola,
garraway1,garraway2,garraway4}. Therefore, our exact solution in
this regime, could be exploitable to describe the dynamics of two
interacting qubits put inside two different optical cavities.

\section{Discussion and conclusive remarks}
The problem of the proper description of open quantum systems is
still far from having a complete and general solution. In
particular, while the features of the Markovian dissipative dynamics
are all well established and accepted, a lot of work has still to be
done to claim general statements on the validity of the different
possible non-Markovian approaches available in the literature.

In this paper we concentrated on the study of two popular methods
aimed at describing non-Markovian dynamics: the Nakajima-Zwanzig and
the time-convolutionless master equation approaches. Exploiting an
exact solution for the dissipative dynamics of two coupled qubits
interacting with independent reservoirs, we have shown that the TCL
approach reproduces all the features of the non-Markovian dynamics
for a range of parameters much wider than the one in which the NZ
equation gives results which are physically reasonable, since the
latter approach may violate the positivity condition on the density
matrix already for reservoir correlations which are not very strong.

The discrimination of the best master equation approach for the
problem under study is a very important issue, because the problem
of the dissipative dynamics of two interacting qubits has been given
a lot of attention in recent years, especially from the point of
view of the entanglement dynamics \cite{quiroga,sinaysky,Tanas,tanas
2008,grifoni,storcz}. So, once we have shown that for some
parameters the TCL master equations provides a very good description
of the non-Markovian dynamics, one may use the same approach for the
description of temperature effects on the system dynamics, since we
do not have at our disposal an exact solution of the dynamics at
general reservoir temperatures.

We finally note that our exact solution can moreover be exploited in
order to study the zero temperature entanglement dynamics in a range
of parameters in which none of the two master equation approaches
can be used: in this way one could extend the studies given in Refs.
\cite{quiroga,scala,sinaysky}, so that one could get the most
general features of the quantum correlations between two qubits
coupled to two independent reservoirs. These points will be the
subject of our future research.

\section{Acknowledgements}
The authors acknowledge stimulating discussions with Prof. A.
Messina. The authors also acknowledge partial support by MIUR
Project N. II04C0E3F3. M.S. acknowledges financial support by the
European Commission project EMALI.

\end{document}